# Revealing timescale-dependent oxygen vacancy distributions in solid oxide fuel cell electrodes using frequency-resolved X-ray absorption (FR-XAS)

Brian Gerwe[1,†], Keita Mizuno[2], Oki Sekizawa[3], Kiyofumi Nitta[3], Koji Amezawa[2], Stuart B Adler[1,‡]

[1] *Department of Chemical Engineering, University of Washington, Seattle, WA, USA*

[2] *Institute of Multidisciplinary Research for Advanced Materials, Tohoku University, Sendai, Japan*

[3] *Japan Synchrotron Radiation Research Institute, Koto, Sayo, Hyogo, Japan*

## Abstract

Development of materials for electrochemical energy conversion requires a deep understanding of the factors governing chemical and physical rates at submicron length scales. Many workers have sought to develop chemically sensitive *in situ* or operando imaging techniques targeting these length scales. However, current methods focus on steady-state or stepwise response. To probe electrode processes both spatially and temporally, we have developed a frequency-resolved implementation of X-ray absorption imaging (FR-XAS) that can measure local electrochemical response in *operando* during a global sinusoidal impedance measurement. Frequency-resolved 1-D images of the oxygen vacancy distribution in a thin film SOFC cathode material ($La_{1-x}Sr_xCoO_{3-\delta}$) reveal, for the first time experimentally, the defect concentrations associated with a Warburg and Gerischer impedance. Analysis of these images allows direct extraction of diffusion and kinetic rate parameters, independent of the global impedance.

## Motivation

Electrochemical impedance spectroscopy (EIS) is often used to probe rates in electrochemical systems as a function of timescale. However, as with any electrochemical measurement, EIS represents the aggregate response of the system, which is a spatial average of rates within the cell. This averaging makes it difficult to quantify rates of individual reaction and transport processes based on EIS features, particularly in samples with inhomogeneous properties.[1,2]

Over the last 30 years[3], workers have developed a variety of chemically sensitive *in situ* and *operando* methods to allow direct measurements of rate phenomena in electrochemical systems with spatial and/or chemical specificity. Lim et al. used operando scanning transmission X-ray microscopy[4,5] to image intraparticle intercalation pathways in Li-ion battery cathodes during dis/charge cycles. Tada et al. directly probed the kinetics of redox processes on Pt catalysts using *operando* time-resolved X-ray absorption fine structure.[6–8] Although these examples involved nonstationary behavior, an inherent limitation of stepwise response is that it simultaneously probes a broad range of timescales.[9,10] In many systems we'd prefer to isolate response at specific frequencies, both to isolate particular rates, as well as to more directly relate *operando* observations to impedance measurements. The present work builds on previous success in measuring the active region of patterned thin film electrodes under steady-state polarization with micro X-ray absorbance spectroscopy (μ-XAS)[11]—by extending these methods into the frequency domain.

†gerweb@uw.edu
‡stuadler@uw.edu

## Patterned Electrode Description

The mixed ionic-electronic conducting (MIEC) cathode $La_{0.6}Sr_{0.4}CoO_{3-\delta}$ (LSC) is often studied as a model electrode for $O_2$ reduction/oxidation in high temperature electrochemical systems and devices.[2] As shown in Figure 1a, oxygen reduction in a porous mixed conducting electrode is thought to occur by a combination of parallel reaction and transport steps. Due to parallel rates, as well as uncertainties in electrode microstructure and material properties, interpretation of the global impedance response in this system is notoriously difficult.[2] In order to mitigate uncertainties in microstructure, as well as provide access to the electrode surface by chemical probes, a number of workers (including the co-authors) have studied electrodes based on patterned thin films.[11–19] As shown in Figure 1b, $O_2$ reduction in a patterned thin film is presumed to occur by the same reaction and transport steps, but with a well-defined geometry that facilitates interpretation of the measured response. It remains an open question whether the surface and bulk *properties* of such films match those of materials in a porous electrode.[17,20,21] Addressing this question is a significant motivation for the present work.

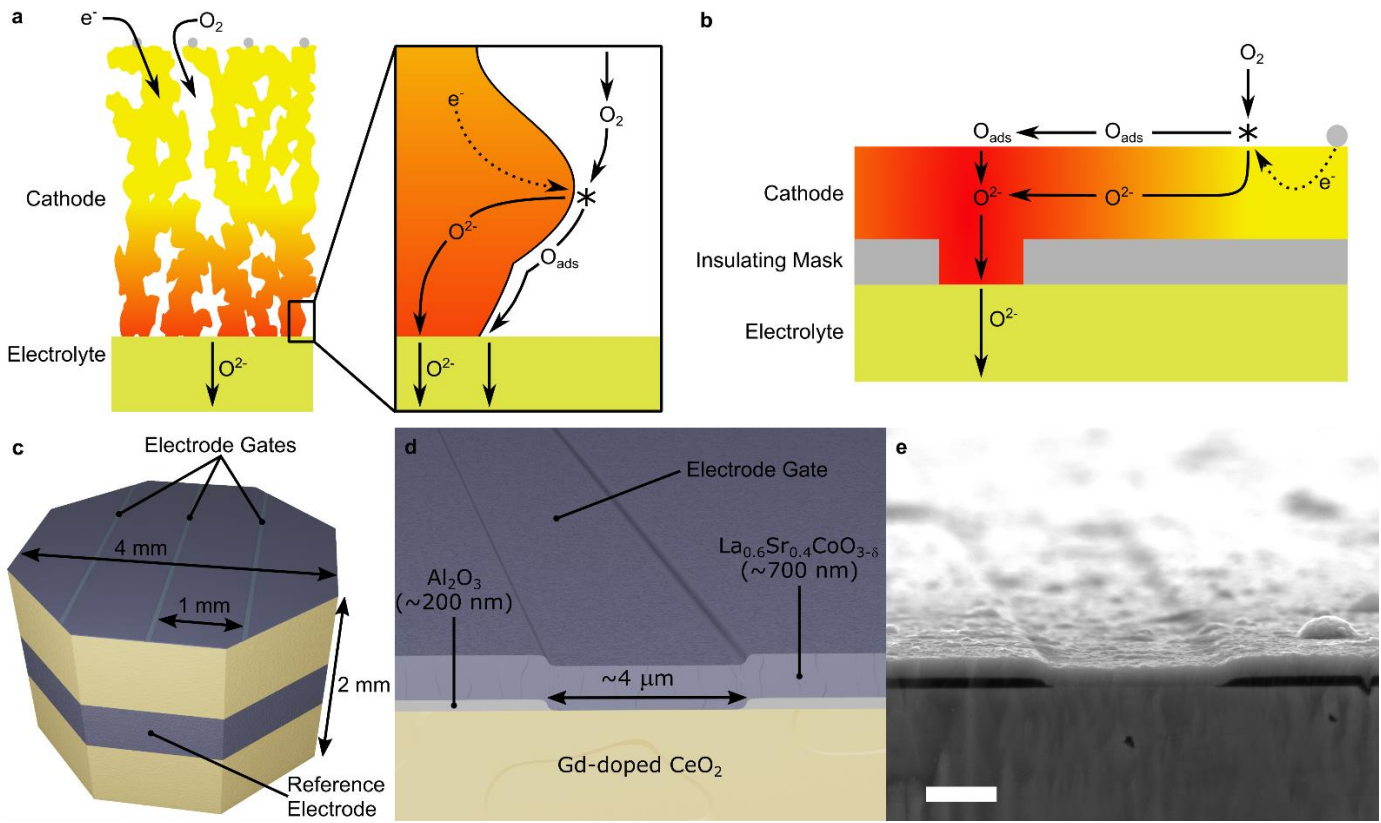

**Figure 1| Laser-deposited patterned LSC thin film sample. a**, Possible pathways for oxygen reduction on a porous mixed ionic-electronic conductor. **b**, Corresponding pathways for oxygen reduction on a patterned thin film electrode. **c**, typical patterned thin film cell geometry. **d**, Typical dimensions and **e**, SEM image of the LSC thin film near an electrode gate. Scale bar, 2 μm. Electrode gates are separated by 1 mm to avoid overlap of vacancy gradients, which may extend up to 200 μm from adjacent gates. Typical film thicknesses are 400–800 nm and 150–300 nm for the electrode and insulating layers, respectively. The electrolyte pellet is approximately 2 mm thick and 16 mm in diameter during film deposition, after which is cut into cuboids about 4 mm in width and height. Finally, the cuboids are trimmed into octagonal prisms to fabricate electrochemical cells for operando characterization. Detailed deposition conditions and fabrication details are included in the Methods section.

As illustrated in Figure 1b, the extended surface area of a porous electrode is emulated in a thin film by limiting electrode/electrolyte contact to a narrow region called the electrode "gate".[11] This is achieved by laser-depositing an insulating layer of amorphous $Al_2O_3$ onto a polished polycrystalline Gd-doped $CeO_2$ electrolyte pellet, lithographically removing portions of specified dimensions, then laser-depositing a thin film of LSC.[11] As shown in Figures 1c and d, electrode gates are formed where gaps in the insulating layer are conformally filled by LSC. The resulting geometry can be thought as a porous electrode laid on its side, with an extended gas-exposed surface accessible to incident and fluoresced X-rays.

## X-ray Spectroscopy as Chemical Probe

Frequency-resolved imaging of local oxidation state in the film was measured using X-ray absorption spectroscopy (XAS).[22,23] Figure 2a shows Co K-edge XAS spectra of an unmasked thin film LSC electrode at 700 °C and a $pO_2$ of 0.10 bar as a function of polarization. The absorption edge near 7720 eV shifts to lower energies under increasingly cathodic (negative) polarization. As shown elsewhere, this (chemical) shift is related to the decrease in average Co valence (increase in oxygen vacancy concentration) upon chemical or electrochemical reduction.[24–27] As such, Co K-edge absorption serves as a measurable proxy for oxygen vacancy concentration. Figure 2b shows the difference spectra of the data in Figure 2a. At energies between the pre-edge features and the isosbestic point (7723 eV), absorption increases under negative polarization. Thus, at a fixed incident energy, reduction/oxidation appears as strong positive/negative changes in absorbance amplitude, with greatest sensitivity near 7720 eV. Figure 2c shows absorbance at a fixed incident energy as a function of distance away from the electrode gate of a patterned thin film electrode under open circuit and negative 150 mV steady polarization. Absorbance is constant across the film under open circuit but increases near the electrode gate. Thus, in agreement with previous work, we can map the active region of the electrode with a spatial resolution of ~1 μm.[11,28]

## Principle of FR-XAS

The core principle of FR-XAS is to spatially-resolve the local changes in film oxidation state described above as a function of frequency during a global sinusoidal potential perturbation. During the steady-periodic modulation, we synchronously record voltage ($V$), current ($I$) and the local X-ray signals—incident flux and fluoresced flux—in a similar manner as nonlinear electrochemical impedance.[29] Details of the experimental setup are included in the Methods section and supplemental materials. Full FR-XAS profiles are constructed by sweeping through a range of positions and perturbation frequencies determined by preliminary steady-state μ-XAS profiles and EIS spectra.

Figure 2d shows the time-domain adsorption $c(t)$ of an unmasked thin film electrode during a sinusoidal voltage perturbation of 0.5 Hz. The top panel shows the voltage perturbation $V(t)$ (closed markers) and corresponding overpotential $\eta(t)$ (open markers) defined by $\eta = V - IR_\Omega$, where $R_\Omega$ is the electrolyte resistance (taken as the high-frequency intercept from impedance measurements). Phase reference lines have been added to mark local troughs in the voltage (dashed cyan line) and overpotential (dot-dashed green line). The bottom panel shows the X-ray absorbance $m(t)$ at 7719.8 eV, defined as the ratio of fluoresced to incident X-ray fluxes. As expected from XAS spectra under steady polarization, absorbance generally decreases as voltage increases; however, as shown by the phase reference lines, absorbance is 180° out of phase with overpotential, not voltage. This alignment suggests that the overpotential is tied primarily to Nernstian shifts in oxidation state in the film, which are opposite in sign from the absorption. In contrast, voltage also contains $IR_W$, which is generally out of phase with overpotential (except under steady polarization), resulting in a significant phase offset. Since in our experiment we measure the phase of the absorption relative to voltage (not $\eta$ directly), we must always include the contribution of ohmic resistance in our image analysis.

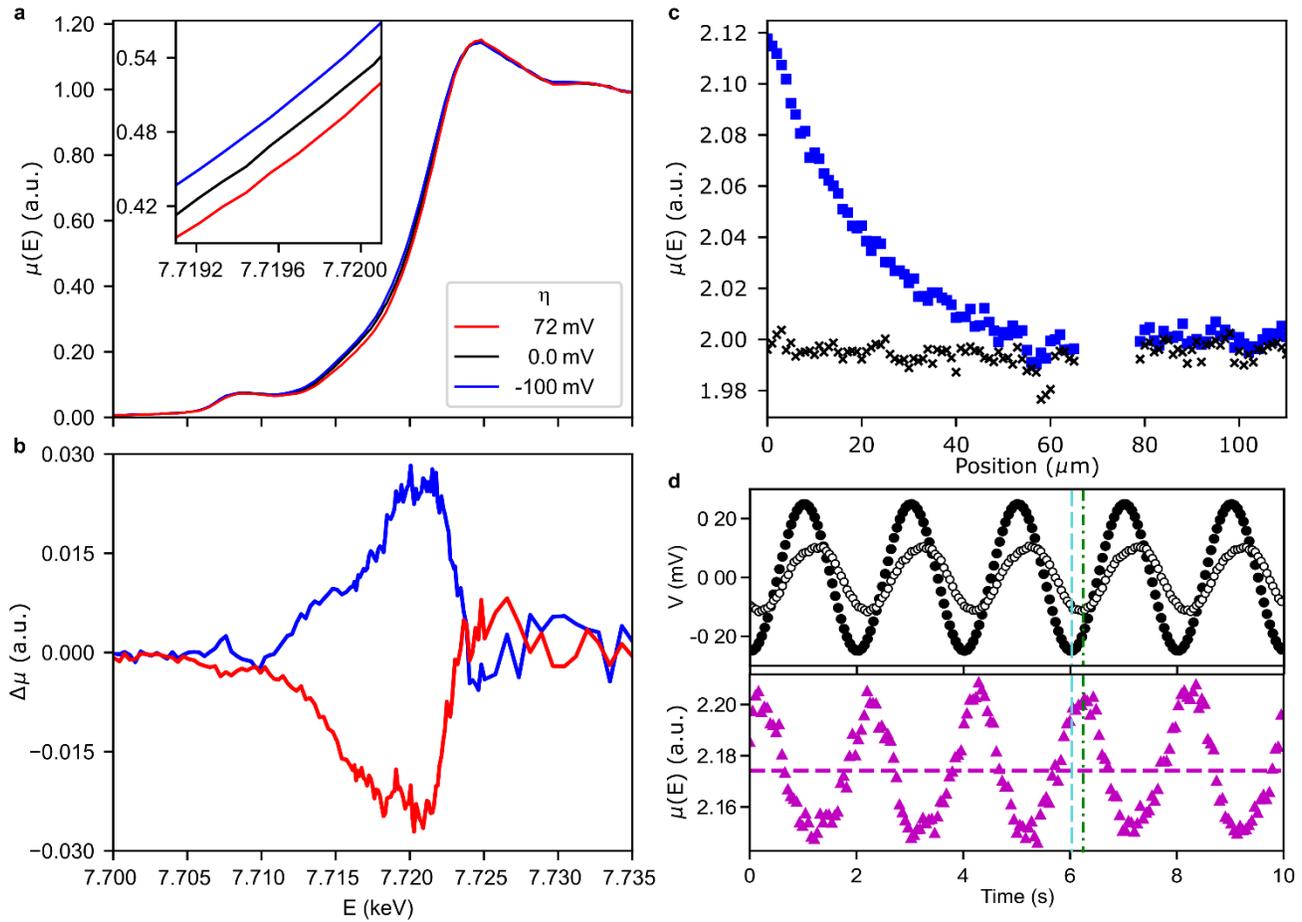

**Figure 2|X-ray absorbance at an electrode gate. a**, Co K-edge XANES spectra at 700 °C and a $p_{O_2}$ of 0.10 bar under open cell voltage, anodic, and cathodic polarization. **b**, Difference spectra of absorbance in panel **a** with the open cell voltage spectrum as the reference. Pre-edge features from 7707 to 7712 eV are attributed to electron state transitions permitted by hybridization of Co 3d and O 2p orbitals, and are mostly insensitive to polarization.[24] **c**, Spatially-resolved absorbance near the electrode gate of a patterned electrode under open circuit (×) and negative 150 mV polarization (■). Data near 70 μm were excluded as outliers due to debris on the film surface. **d**, Time-domain waveforms of voltage (●), overpotential (o), and absorbance (▲) for a 250 mV 0.5 Hz sinusoidal polarization. Vertical dashed cyan and green dash-dotted lines indicate troughs of the voltage and overpotential signals, respectively. Horizontal red dashed line indicates the mean absorbance level.

Although many factors influence the absolute magnitude of μ, we are primarily interested in deviations of μ from its value at equilibrium, since these can be related to vacancy concentration fluctuations.[11] Therefore, we define a relative absorbance, $\chi = \frac{\mu - \mu_o}{\mu_o}$, where $\mu_o$ is the absorbance under equilibrium conditions. Vacancy concentration fluctuations are treated similarly by defining $\psi = \frac{x_v - x_v^o}{x_v^o}$, where $x_v$ is the oxygen vacancy mole fraction, and $x_v^o$ is the value of $x_v$ at equilibrium at the measurement temperature and $p_{O_2}$. Under the conditions and perturbation amplitudes of the experiment, we assume a linear relationship between absorbance and vacancy fluctuations: $\chi = \beta\psi$, where $\beta$ is a measurable coefficient.

For a patterned thin film, the relative absorption $\chi(\tilde{\omega};y,t)$ is function of position ($y$) from the gate as well as time ($t$) and radial frequncy ($\tilde{\omega}$) of the voltage perturbation. However, as with EIS, for small steady-periodic perturbations the absorbance will have form $\chi(\tilde{\omega};y,t) = \frac{1}{2}\overline{\chi}(\tilde{\omega};y)e^{j\tilde{\omega}t} + \frac{1}{2}\overline{\chi}*(\tilde{\omega};y)e^{-j\tilde{\omega}t}$, where $\overline{\chi}(\tilde{\omega};y)$ is the time-Fourier transform of $C$. Thus by fourier transforming the time domain absorbance data, we obtain a complex but *stationary* image profile $\overline{\chi}(\tilde{\omega};y)$ with real and imaginary components that depend on kinetic and transport rates.

The Fourier analysis is performed using an open source Python package called frxas.py[30], and roughly follows the procedure in Ref. 29 for analyzing nonlinear EIS. Briefly, time-domain data are apodized by a Gaussian window function before applying a complex fast Fourier transform. Real and imaginary components of the signal at the oscillation frequency are extracted by fitting the resulting frequency spectrum to the corresponding Gaussian and Dawson basis functions. The phase of each signal relative to the voltage perturbation (maximum cathodic voltage at t=0) is established by finding the absolute voltage phase angle $\left(\varphi_1 = \text{atan2}(\text{Im}[\overline{\mathbf{V}}], \text{Re}[\overline{\mathbf{V}}])\right)$ and multiplying a zeroth-order phase correction $\exp(j(\pi-\varphi_1))$ to all signals. Finally, FR-XAS profiles are generated by plotting the real and imaginary components of $\overline{C}$ as a function of distance $y$ from the electrode gate.

## FR-XAS Results

Figure 3 shows FR-XAS profiles at 700 °C under several $pO_2$ for 150 mV amplitude voltage perturbations of various frequencies ($\tilde{f}$). Profiles for $\tilde{f}$ = 0.00 Hz (steady state) were collected separately under DC polarization before FR-XAS measurements. Since the film experiences both cathodic and anodic polarization during oscillations, the steady-state profiles were similarly determined by measuring profiles under positive and negative bias, inverting the negative profile, and averaging. At low oscillation frequencies (e.g. Figure 3b, 0.25 Hz) the region of nonzero $\overline{C}$ becomes more confined to the gate, and $\text{Im}[\overline{C}]$ becomes nonzero, indicating a position-dependent phase shift in the vacancy concentration. With increasing frequency, $\text{Re}[\overline{C}]$ and $\text{Im}[\overline{C}]$ at the gate decrease, which we attribute to the overpotential amplitude lowering from ohmic drop in the electrolyte.

At moderate and high frequencies (e.g. Figure 3b, 1 and 5 Hz) perturbations in vacancy concentration become further confined to the gate, and exhibit a self-similar profile with a prominent region of negative $\text{Re}[\overline{C}]$. At these higher frequencies the width of the active region decreases with increasing frequency, scaling approximately as $1/\sqrt{2\pi\tilde{f}}$. This profile and scaling closely matches the theoretical predictions of the Warburg impedance for semi-infinite diffusion.[2,31] To the best of the author's knowledge, Figure 3 shows the first direct measurements of a Warburg concentration profile during an impedance measurement.

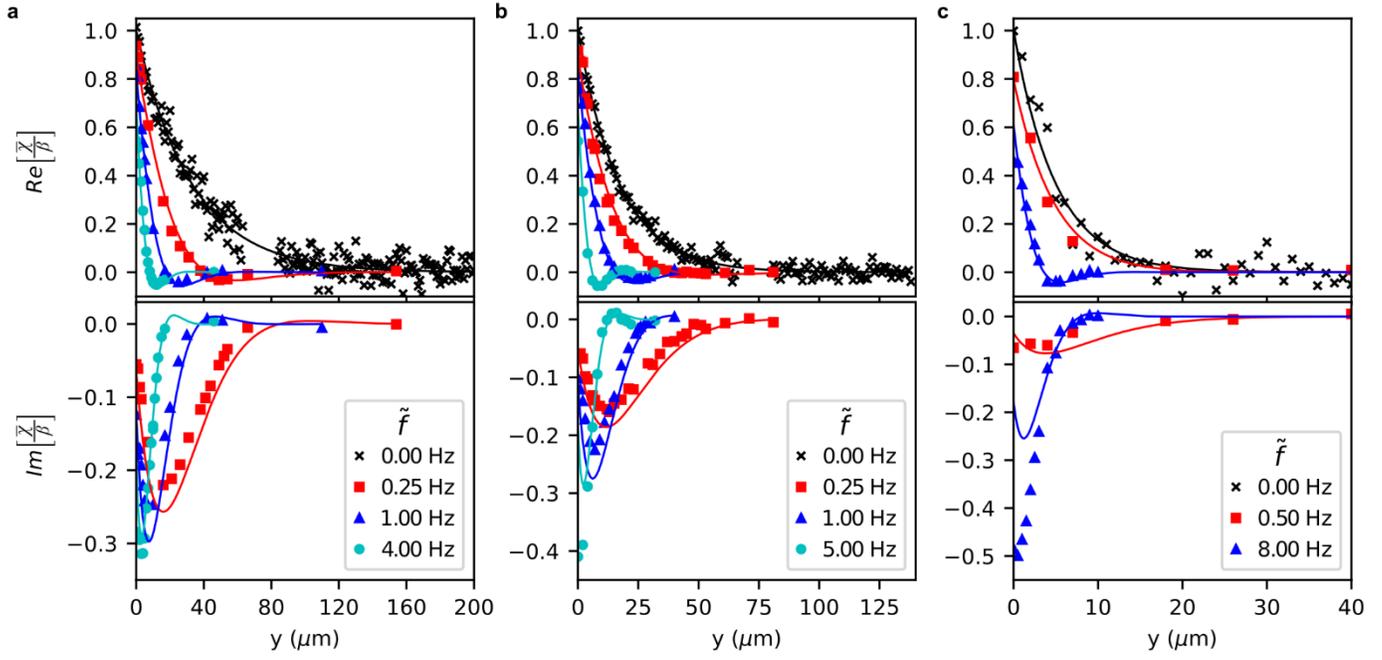

**Figure 3 | FR-XAS profiles measured at 700 °C under various $pO_2$.** **a**, 0.01 bar. **b**, 0.10 bar. **c**, 1.00 bar. Steady state profiles before FR-XAS measurements using a separate fluoresence detector. Lines indicate best fit to equation (1). Profiles are normalized by $\beta$, the empirical coefficient relating absorbance and oxygen vacancy displacements.

## One-Dimensional Model

The data in Figure 3 were analyzed using a 1-D model following a framework used by Adler, Lane and Steele for porous electrodes.[32] The primary assumption of this model is that vacancy gradients extend over distances much larger than the film thickness (600 nm), such that gradients across the film thickness are negligible. The model considers a dynamic material balance of oxygen vacancies in the mixed conductor with bulk transport of vacancies, and oxygen reduction on the surface $\left(1/2 O_2 + 2e' + V_O^{\bullet\bullet} \rightleftarrows O_O^X\right)$ treated as a homogeneous reaction within a 1-D continuum. We assume oxygen transfer to/from the electrolyte is blocked by the insulating layer everywhere except at the gate, which comprises one boundary condition for the 1-D model. At this boundary, the electrolyte/electrode interface is assumed to be equilibrated, such that defect concentration at the gate is controlled by overpotential ($\eta$), creating a driving force for oxygen exchange and vacancy diffusion propagating away from the gate.

Since we expect the profiles to be steady-periodic, the governing equations for this model are best solved by transforming into the frequency domain, allowing simple separation of the time and spatial components[29] (see supplemental materials). For a small amplitude voltage perturbation of frequency $\tilde{\omega}/2\pi$, the time-Fourier transform of the vacancy concentration profile predicted by the model is given by:

$$\overline{\psi}(\tilde{\omega}) = \frac{e^{-y/l_\delta \sqrt{1+j\tilde{\omega}t_G}}}{\left(1+\gamma\sqrt{1+j\tilde{\omega}t_G}\right)} \qquad (1)$$

where $l_\delta$ (utilization length) is a characteristic length proportional to the size of the active region at steady state, $t_G$ (Gerischer time constant) is a characteristic timescale for diffusion over distance $l_\delta$, and $\gamma$ is the ratio of electrolyte resistance to electrode resistance. The quantities $l_\delta$ and $t_G$ are related to thermodynamic, transport, and kinetic properties by:

$$l_\delta = \sqrt{\frac{c_O x_v^o D_v L}{4\mathfrak{R}_O^o}} \; ; \qquad t_G = \frac{c_O x_v^o L}{4 A_o \mathfrak{R}_O^o}, \qquad (2\text{ a,b})$$

where $c_O$ is the molar concentration of oxygen lattice sites, $x_v^o$ is the mole fraction of oxygen vacancies at equilibrium with the surrounding gas, $A_o = -\frac{1}{2}\frac{\partial \ln f_{O_2}}{\partial \ln x_v}$ is the thermodynamic factor relating changes in composition to change changes in oxygen chemical potential via $pO_2$ or voltage, $D_v$ is the oxygen vacancy diffusion coefficient, $\mathfrak{R}_O^o$ is the equilibrium molar exchange rate of $O_2$ between the mixed conductor surface and gas (as defined in Ref. 33), and $L$ is the thin film electrode thickness.

In the limit of low frequency ($\tilde{\omega} t_G \ll 1$), Equation 1 reduces to a purely real exponential decay over distance $l_\delta$.

$$\psi = \frac{1}{1+\gamma} e^{-y/l_\delta} \qquad (3)$$

This profile reflects $O_2$ reduction becoming co-limited by diffusion and kinetics, where the size of the active region is determined by both processes. The lack of an imaginary component indicates that lateral diffusion is fast enough to reestablish a quasi-steady profile as the defect concentration at the gate is modulated with AC polarization. In the limit of high frequency ($\tilde{\omega} t_G \gg 1$), the profile reduces to a Warburg profile:

$$\psi \approx \frac{e^{-y\sqrt{\frac{j\tilde{\omega}}{A_o D_v}}}}{\left(1+\gamma\sqrt{j\tilde{\omega} t_G}\right)} \qquad (4)$$

Corresponding to a purely diffusion-limited response.[2,31] In the Warburg limit, we expect the profile to be independent of the surface exchange rate, allowing us to isolate and quantify the diffusion rate.

The term containing $\gamma$ in the denominator of Equation 1 represents the impact of Ohmic drop in the electrolyte. Assuming no resistance to ion transfer between electrolyte and mixed conductor, concentration fluctuations at the gate will always be in-phase with the overpotential ($\eta$) through the Nernst equation. Thus if $\gamma = 0$ (no ohmic resistance), concentration fluctuations at the gate will be in-phase with applied voltage $\eta = V$. However, with nonzero $\gamma$, $\eta = V - IR_\Omega$, and thus $\psi$ at the gate decreases in magnitude and contains a complex component that shares phase with the current. This phase shift is evident in Figure 3, which shows a nonzero imaginary signal in the FR-XAS profiles relative to $V$ even at the gate.

## Model Fitting and Parameter Extraction

To fit the model described above to the data in Figure 3, Equation 1 must be multiplied by $\beta$, the coefficient (described previously) linking vacancy and absorbance displacements. Unfortunately, we found $\beta$ to be highly variable among samples, so ignored its value in parameter extraction. Rather, the absorbance amplitude data in Figure 3 is normalized to its magnitude at the gate, and the value of $\beta$ is taken to be unique for each profile. For the remaining fit, the values of $l_\delta$, $t_G$, and $\gamma$ (which represent characteristics of the cell) are assumed to be independent of frequency, and thus *shared* among all profiles at a given $pO_2$. The values of $l_\delta$ are extracted solely from the steady-state profiles (zero frequency) by fitting to an exponential decay with a baseline correction. The profiles at non-zero frequency are then used to fit values of $t_G$ and $\gamma$.

The lines in Figure 3 show the resulting fits. The model appears to accurately capture both the spatial and frequency dependence of the defect profile, with largest discrepancies near the gate, which is particularly pronounced at higher $pO_2$. Based on ongoing modeling work using finite element analysis, we currently believe this discrepancy arises primarily from 2-D diffusion of oxygen ions in the gate vicinity. Table 1 shows the modeling parameters extracted from these fits, as well as the vacancy diffusion and surface exchange rate coefficients calculated by inverting Eqns. 2a and 2b. For this calculation, the values of $A_o$ and $x_v^o$ were determined from independent measurements of chemical capacitance on unmasked thin films fabricated using identical laser-deposition conditions[34], using the method of Kawada et al.[20]

As expected, both $l_\delta$ and $t_G$ decrease with $pO_2$, reflecting the dominant role of decreased vacancy concentration. The calculated values in Table I suggest that $D_v$ is only weakly $pO_2$-dependent, whereas $\mathfrak{R}_O^\circ$ varies by an order of magnitude over the measured conditions. These trends agree well with previous work by Lu and others on porous LSC electrodes of the same composition.[33,35] However, the absolute values of $D_v$ and $\mathfrak{R}_O^\circ$ values reported here are approximately one order of magnitude higher and lower, respectively, than the values for a porous electrode of the same composition[25].

One possible explanation for these discrepancies is a true difference in properties between a thin film and a porous material. It is well-known that thermodynamic properties of constrained thin films differ dramatically from bulk materials of the same cation composition.[16–18,36,37] Also, unlike sintered particles, the nanostructure of thin films often involves closely packed columns, where boundaries between columns could promote fast diffusion pathways.[38,39] Finally, surface exchange is highly sensitive to Sr enrichment and/or Sr-rich precipitates on the surface, which is known to dramatically impact surface exchange of $O_2$.[16,40] Although enrichment has been documented for both bulk and thin film samples, minor differences could explain the gap in apparent surface exchange rates.[18,41–43]

However, the anti-correlated discrepancies in $\mathfrak{R}_O^o$ and $D_v$ relative to porous electrodes may not be a coincidence. During their analysis, Lu et al did not measure the vacancy concentration profile – they had to *infer* the size of the active region $l_\delta$ based on fitting of the impedance data. Any errors in their inference of $l_\delta$ would have anti-correlated impacts on the values of $\mathfrak{R}_O^o$ and $D_v$ extracted from the data. In contrast, the study here involves *direct extraction of* both $l_\delta$ and $t_G$ as characteristics of the measured concertation profile. Thus, we have a much higher level of confidence in the relative rates of diffusion and kinetics extracted from the image data. Perhaps further studies of porous electrodes (using the techniques described in this paper) could resolve whether there is a true difference in kinetic and transport parameters for porous materials vs thin films.

**Table 1 | Best fit parameters and calculated quantities from FR-XAS** Utilization lengths and Gerischer time constants from best fits of equation (1) to profiles in **Figure 3**, including 95% confidence intervals from the regression analysis. Vacancy diffusion and equilibrium exchange rate coefficients are calculated using best fit parameter values. Equilibrium oxygen vacancy mol fractions and thermodynamic factors estimated from impedance measurements on a laser deposited LSC thin film at 700 °C under a wide range of $pO_2$.

|  | $p_{O_2}$ (bar) | | |
| --- | --- | --- | --- |
|  | 0.01 | 0.10 | 1.00 |
| $l_\delta$ (μm) | $32.8 \pm 1.03$ | $17.1 \pm 0.844$ | $4.98 \pm 0.589$ |
| $t_G$ (s) | $2.34 \pm 0.168$ | $0.946 \pm 0.0891$ | $0.145 \pm 0.0314$ |
| $D_v$ (cm²/s) | $1.70 \times 10^{-6}$ | $1.49 \times 10^{-6}$ | $1.02 \times 10^{-6}$ |
| $\Re_O^\circ$ (mol/ cm² s) | $4.47 \times 10^{-9}$ | $9.30 \times 10^{-9}$ | $4.36 \times 10^{-8}$ |
| $x_v^o$ | $2.0 \times 10^{-2}$ | $1.3 \times 10^{-2}$ | $7.5 \times 10^{-3}$ |
| $A_o$ | 2.70 | 2.09 | 1.67 |

## Comparing 1-dimensional impedance model to measurements

Based on the rate parameters in Table 1, we should, in principle, be able to predict *a priori* the impedance response of the electrode using our 1-D model without fitting the impedance data. This model predicts a total impedance given by

$$Z = R_\Omega + \frac{R_G}{\sqrt{1 + j\tilde{\omega} t_G}}; \; R_G = \frac{RT}{8F^2} \frac{A_o t_G}{c_O l_d L x_v^o W} \tag{5}$$

where $T$ is temperature, $R$ is the ideal gas constant, $F$ is Faraday's constant, $W$ is the sum of electrode gate lengths, and $R_W$ is the electrolyte resistance.

Figure 4a compares the measured impedance spectra of the patterned electrode at 700 °C before FR-XAS measurements to the predictions from Equation 5 using parameters from Table 1 (with $R_W$ taken from the measured high frequency intercept of the impedance). Although the measured impedance follows a similar overall trend with $pO_2$, it is approximately half the magnitude of the predicted impedance, with characteristic frequency (point of maximum imaginary amplitude) about 3~4 times higher than predicted. The shape of the measured impedance arc (on a Nyquist plot) at high frequency is also inconsistent with the 45 deg. Warburg diffusion limit predicted by the model, and contains additional small features at high and low frequency.

We currently hypothesize that these inconsistencies arise from three principle sources: 1) break-down of the 1-D diffusion approximation for gradients near the gates, 2) leak of current through the insulating mask, and 3) frequency-dependent current constriction in the electrolyte near the gate. These hypotheses are supported by our separate characterization of the mask properties and ongoing 2-D finite element modeling, which will be released in a subsequent publication. However, in the meantime we have also developed a simple equivalent circuit model (described in greater detail in the supplemental materials) to help describe these contributions. As shown in Figure 4b, when this model (dash-dot line) is fit to the data, all parameters fall within a factor of 2 of our estimates of known contributions. This model also accurately predicts the shape of the impedance at high frequency, as well as the dispersion at low frequencies in the 0.10 and 1.0 bar measurements. This agreement is remarkable given that the 1-D images in Fig. 3 (from which the model parameters are extracted) represents a tiny fraction the much larger, possibly inhomogeneous, sample.

Note that in the absence of the rate parameters extracted from the images in Figure 3, it is <u>unlikely</u> we could have conceived of all the additional factors impacting the impedance in Figure 4. Even if we had conceived of them, it would have been difficult to draw any strong conclusions about the material properties based on fits to the impedance data alone, due to the very large number of adjustable or uncertain parameters. What makes any of this analysis possible is the *direct observation* of electrode concentration fluctuations and utilization length as functions of $pO_2$ and frequency. For these reasons we believe FR-XAS, and its ability to probe spatiotemporal correlations in *operando*, holds great promise for distinguishing rates in solid state electrochemistry.

## Outlook

We have introduced frequency-resolved X-ray absorption spectroscopy and presented the first 1-D images of concentration profiles associated with the Gerischer impedance. By analyzing these images with a 1-D model, we extracted diffusion and kinetic parameters for the oxygen reduction reaction in a mixed ionic-electronic conducting cathode. Due to other process occurring in the sample over similar timescales, this would not be possible using conventional electrochemical techniques alone. This demonstrates the utility of chemically sensitive, frequency-resolved techniques for gaining a deeper understanding of the chemical and physical phenomena governing materials used for electrochemical energy conversion. In principle, material systems used for other electrochemical devices may be studied by adapting suitable imaging modalities to a frequency-resolved approach.

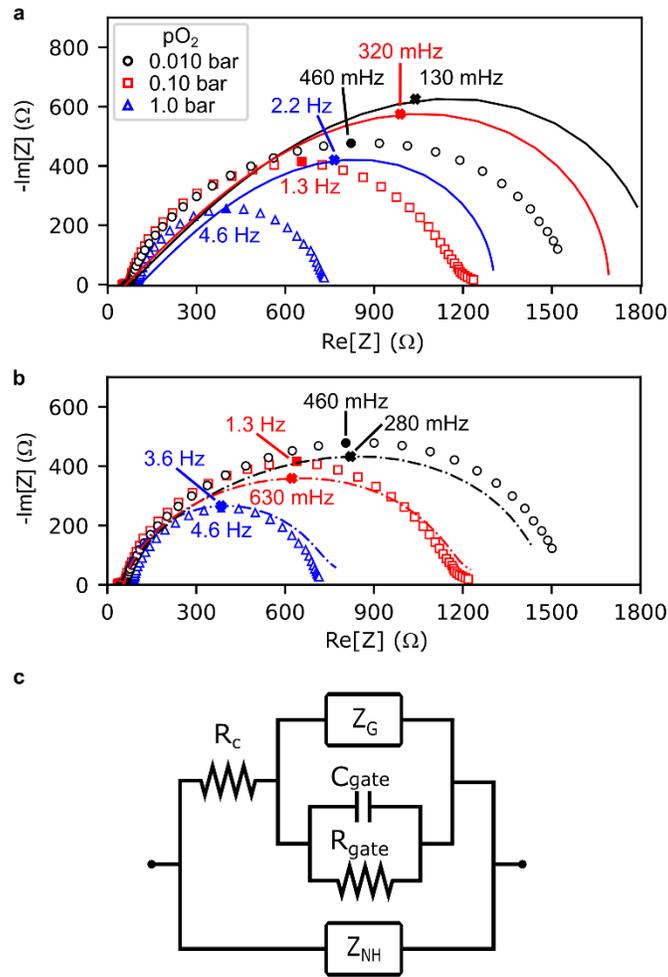

**Figure 4 | Patterned thin film impedance. a**, Measured impedance spectra at 700 °C after subtracting estimated electrolyte resistance (markers) compared to predicted Gerischer impedance (lines) using parameters in Table I. **b**, Measured impedance spectra (markers) compared to impedance predicted by the proposed equivalent circuit model. **c,** Proposed equivalent circuit model including contributions in parallel with the Gerischer impedance ($Z_G$): oxygen activity in the electrode above the gates ($R_{gate}$, $C_{gate}$), current constriction into the gates ($R_c$), and impedance of the insulating layer ($Z_{NH}$).

# Methods

## Sample Synthesis

The patterned thin film sample was prepared using pulsed laser deposition (PLD) on a dense $Gd_{0.1}Ce_{0.9}O_{1.95}$ (GDC) pellet electrolyte. The electrolyte was prepared by uniaxially pressing ~3 g of GDC powder (Shin-Etsu Chemical Co., Ltd) in a 20 mm diameter die. The resulting pellet was hydrostatically pressed at 200 MPa followed by sintering at 1550 °C for 5 hours. Upon sintering, the pellet shrank to ~16 mm in diameter with a 97% relative density. One side of the pellet was mirror-polished to a surface roughness of 2~3 nm (Disco, Inc.). The other side was painted with a 60/40 wt% LSC/Pt (AGC Seimi Chemical Co., Ltd./ Tanaka Kikinzoku Kogyo Co., Ltd., TR-7907) paste as a counter electrode and sintered at 950 °C for 3 hours. Image reversal photoresist (Clariant International Ltd, AZ 5214E) was applied to the polished side, exposed to masked UV light and developed with 2.38% TMAH (Tama Chemicals Co., Ltd.) resulting in ~2 µm wide strips separated by ~1 mm. An insulating layer of $Al_2O_3$ (Nikkato Co., SSA-S) was then deposited with PLD with the following deposition parameters: 180 mW quadruple Nd:YAG laser with λ=266 nm, 10 Hz pulse rate at room temperature under a $pO_2$ of $10^{-5}$ bar for 2 hours. Acetone was used to dissolve the photoresist strips and lift-off overlaying portions of the $Al_2O_3$ film before annealing in air at 900 °C for 3 hours. A 13 mm diameter LSC electrode was then deposited by PLD at 800 °C under a $pO_2$ of $10^{-5}$ bar for 1.5 hours and annealed for 4 hours at 800 °C under a $pO_2$ of 1 bar. The resulting sample had an LSC film ~600 nm thick on top of a ~200 nm thick $Al_2O_3$ film with ~4 μm wide regions of electrode/electrolyte double phase boundary (DPB) separated by 1 mm where the $Al_2O_3$ layer was lifted off. This sample was cut with a diamond saw into several 4 mm X 4 mm X 2 mm cubes to prepare electrochemical cells for FR-XAS measurements.

## Electrochemical Cell and Sample Holder

Electrochemical cells were fabricated by placing a patterned thin film sample onto a Pt mesh and wire feed through a dual-bore alumina tube. A porous Pt electrode painted around the perimeter of the electrolyte served as the reference electrode. A Pt/10% Rh wire was spot welded to the Pt reference electrode wire to create an S-type thermocouple; the positive lead of this thermocouple was left electrically floating during all electrochemical and X-ray measurements. Another Pt wire with Pt mesh was placed onto the LSC thin film edge as the working electrode current collector. A C-shaped alumina plate was cemented (Inorganic Adhesive D, Aron Ceramics) to the end of a single-bore alumina tube, attached over the smaller alumina tube (to maintain contact between the Pt mesh current collector and LSC electrode) and cemented in place. Resistance heating wire was coiled around the end of the larger alumina tube and cemented (Inorganic Adhesive D, Aron Ceramics) in place as a heating element. This assembly was then fed into a slip-on flange with alumina-insulated feedthroughs for the heating element leads and a gas outlet. Finally, the flange was connected to a custom-fabricated jacketed chamber with a Kapton film window to control the gas environment while maintaining X-ray transparency.

## X-ray Measurements

All X-ray measurements were performed using a scanning X-ray microspectroscopy system at BL37XU of SPring-8, Japan . The incident X-ray energy was adjusted by a Si(111) double crystal monochromator and the higher-order harmonic X-ray were rejected by a Rh-coated mirrors with glancing angle of 4 mrad. The incident X-ray was focused to 0.6 x 0.9 μm by a pair of Karkpatrick-Baez mirrors. Co K-edge X-ray absorption near edge structure (XANES) was performed using incident X-ray energies ranging from 7.6 to 7.85 keV. Incident X-ray intensity was measured by an ionization chamber(S-1196A1, OHYO KOKEN KOGYO Co., Ltd.) with a current amplifier (428, Tektronix, Inc. (formerly Keithley Instruments Inc.)) and fluoresced X-ray intensity was measured by two four-element silicon drift detectors (Vortex-ME4, Hitachi High-Tech Science America Inc.) with digital signal processor (Xspress3, Quantum Detectors Ltd.). XANES spectra at a DPB on the sample were collected at 700 °C under $p_{O2}$'s of 1, 0.1, and 0.01 bar, controlled by premixed $O_2$-He gas cylinders. XANES spectra were also collected under cathodic and anodic polarization—typically +/- 150 mV—using a VersaSTAT 3 potentiostat to evaluate the absorption edge shift as a function of the film oxidation state. *Operando* micro XAS (μ-XAS) measurements were performed at a fixed incident X-ray energy of 7719.8 eV under open circuit conditions and cathodic polarization to establish the steady-state oxygen potential distribution.

# Supplemental Information

## Experimental Apparatus

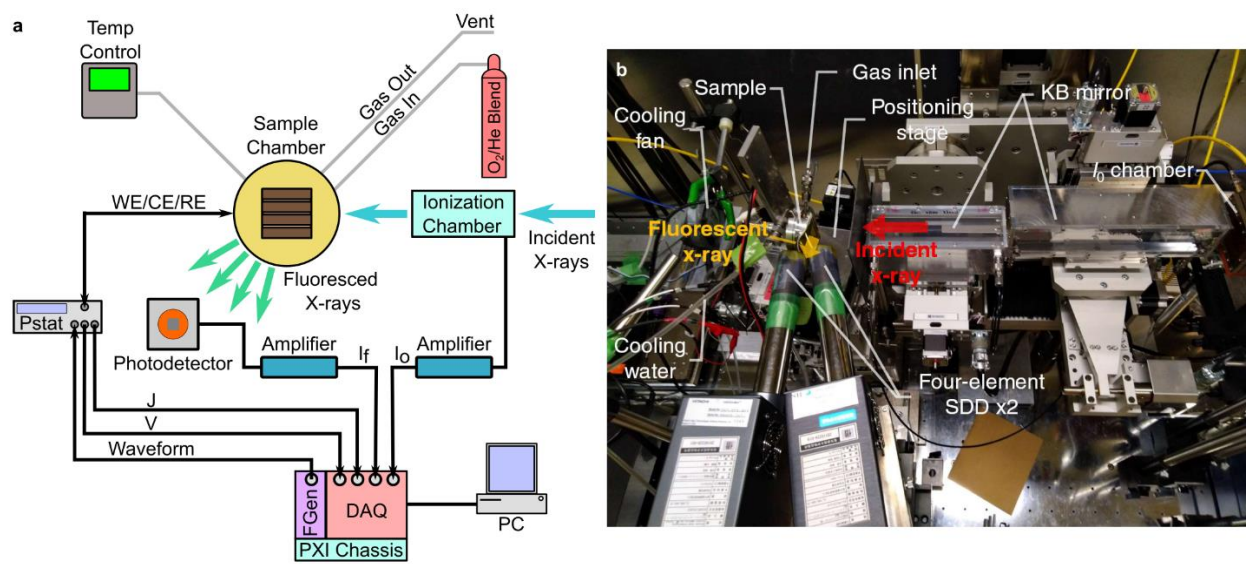

**Figure S1** | a, schematic of experimental setup for FR-XAS. b, Overview picture of the experimental setup for XANES and *operando* μ-XAS measurements. The potentiostat used during these measurements is not pictured. Note that XANES and μ-XAS measurements use different fluorescent X-ray detectors than FR-XAS measurements.

A schematic of the experimental setup is shown in Figure S1a. Control of the sample temperature and gas environment are achieved by mounting an electrochemical cell of the patterned electrode film in a custom sample chamber as detailed in the Methods section of the main text. The incident X-ray intensity was measured by an ionization chamber and the fluorescent X-ray intensity was measured by a Si-PIN photodiode (S-2500, OHYO KOKEN KOGYO Co., Ltd.); both signals were boosted to measurable levels with a low noise current amplifier (DLPCA-200, FEMTO Messtechnik GmbH). The measurement position of the sample was controlled in the x, y, and z axes by actuators moving the entire sample chamber relative to the incident X-ray beam. The incident beam was focused to 910 nm x 250 nm by a pair of Kirkpatrick-Baez mirrors.

Sample polarization is dynamically controlled by a potentiostat (Pstat; Solartron 1287) driven by the output signal from an arbitrary waveform generator (FGen; NI PXIe-5451). A data acquisition device (DAQ; NI PXIe-6366) controlled by a PC collects the incident and fluoresced X-ray signals synchronously with the voltage (V) and current (J) signals from the potentiostat. Synchronicity between data signals is crucial to maintain phase coherence required for data analysis and interpretation.

## One-Dimensional Model Derivation

Eq. 1 is based on a model originally proposed by Adler, Lane and Steele[1], adapted to $La_{0.6}Sr_{0.4}CoO_{3-\delta}$ (LSC) for our patterned thin film electrode geometry. The central hypothesis of this model is that bulk oxygen vacancy transport across the film thickness is equilibrated, and thus accumulation of oxygen vacancies in the film can be expressed as a 1-D differential balance involving lateral bulk transport and reaction at the gas-exposed surface:

$$c_O \frac{\partial x_v}{\partial t} = -\frac{\partial N_v}{\partial y} - \frac{2r_{O_2}}{L} \qquad (S1.1)$$

where $x_v(y,t)$ is the mole fraction of oxygen vacancies, $t$ is time, $y$ is lateral distance from the gate, $N_v$ is the area-specific transport flux of bulk oxygen vacancies in the y direction, $r_{O_2}$ is the local rate of $O_2$ reduction (with units moles $O_2$ cm$^{-2}$ s$^{-1}$), $c_O$ is the site concentration of the oxygen sublattice (a constant), and $L$ is the film thickness.

The model assumes vacancy transport occurs only through bulk diffusion in the mixed conductor, following Fick's Law for a thermodynamic driving force:

$$N_v = -D_v c_O A(x, x_v) \frac{\partial x_v}{\partial y} \qquad (S1.2)$$

where $D_v$ is the vacancy diffusion coefficient. $A(x, x_v) = -\frac{1}{2} \frac{\partial \ln pO_2}{\partial \ln x_v}$ is a thermodynamic factor relating vacancy concentration to oxygen chemical potential in the film. As described previously,[2,3] for LSC this factor has been found by Mizsuzaki[3] and Lankhorst[2] to closely obey:[4,5]

$$A(x, x_v) = 1 + \frac{b(x)x_v}{RT} \qquad (S1.3)$$

where $x_v = \delta/3$ is the mole fraction of oxygen vacancies and $b(x)$ is a parameter sensitive to the Sr doping level in La$_{1-x}$Sr$_x$O$_{3-\delta}$, which is thought to capture the enthalpic contributions to free energy from changes in oxygen nonstoichiometry (~897 kJ mol$^{-1}$ for the film studied here[6]).

Previous work suggests the $O_2$ reduction on LSC is first order in pO2 and 2$^{nd}$ order in oxygen vacancy concentration over a wide range of conditions.[2,3,7–9] This assumption leads to the following rate-law proposed by Kreller et al:

$$r_{O_2} = \mathfrak{R}_O^o \left[ \left( \frac{x_v}{x_v^o} \right)^2 - e^{-2\frac{b}{RT}(x_v - x_v^o)} \right], \qquad (S1.4)$$

where $x_v^0(p_{O_2})$ is the equilibrium vacancy mole fraction for a given pO2 in the surrounding gas, and $\hat{A}_O^o(p_{O_2}) = k p_{O_2}(x_v^0)^2$ is the equilibrium surface exchange rate[8]. This rate-law agrees well with reports of the exchange rate as a function of $pO_2$ on bulk samples as well as the nonlinear electrochemical impedance response of thin films across a wide range of temperature and $pO_2$.[2,7,8,10,11]

Substituting (S1.2), (S1.3) and (S1.4) into (S1.1) yields:

$$c_O \frac{\partial x_v}{\partial t} = -\frac{\partial}{\partial y}\left( -D_v c_O \left( A_o + (A_o - 1)\frac{x_v - x_v^o}{x_v^o} \right) \frac{\partial x_v}{\partial y} \right) - \frac{2\mathfrak{R}_O^o}{L}\left( \left( \frac{x_v}{x_v^o} \right)^2 - e^{-2\frac{b}{RT}(x_v - x_v^o)} \right) \qquad (S1.5)$$

where $A_o(p_{O_2})$ is the thermodynamic factor when the film is equilibrated with the $pO_2$ of the surrounding gas.

The solution to Eq. S1.6 requires two spatial boundary conditions. Since the active region for oxygen reduction (maximum of ~200 μm) is much smaller than the physical distance between adjacent gates (~1 mm), we apply a semi-infinite boundary condition that requires the oxygen nonstoichiometry far from the gate to remain in equilibrium with the gaseous $pO_2$:

$$x_v\big|_{y\to\infty} = x_v^o \tag{S1.6}$$

At the gate ($y = 0$), the lateral flow of vacancies must equal the net ionic current to the gate from the electrolyte. Both these flows are functions of the applied voltage, leading to a compound boundary condition involving vacancy flux and vacancy concentration. One component of this condition is Faraday's law:

$$I/2 = -2F N_v\big|_{y=0} LW, \tag{S1.7}$$

where $I$ is the total anodic current (half of which is assumed to enter the electrolyte from each of the two adjoining segments of the film), $F$ is Faraday's constant, $L$ is film thickness and $W$ is the sum of the electrode gate lengths (approximately 3 mm). The second component is the assumption that ion exchange at the electrode/electrolyte interface remains in equilibrium (as corroborated by the data in Fig. 2d). This assumption implies a Nernstian overpotential given by:

$$\eta = V - IR_W = \frac{RT}{4F} \ln\left[\frac{f_{O_2}\big|_{y=0}}{p_{O_2}}\right], \tag{S1.8}$$

where $f_{O_2}\big|_{y=0}$ is oxygen fugacity in the mixed-conductor at the gate, $R_\Omega$ is electrolyte resistance, and $V$ is cell voltage measured with respect to a reference electrode at the same $p_{O2}$ as the thin film electrode, located on the side of the electrolyte pellet far from the gate.

Applying the Mizusaki-Lankhorst thermodynamic model adopted by Kreller et al[3], oxygen fugacity can be re-expressed in terms of vacancy mole fraction, while vacancy flux can be related to current through Faraday's law (above). These substitutions lead to a compound boundary condition relating $x_v$ and $N_v$ at y = 0:

$$V = -\frac{RT}{2F}\left(\ln\frac{x_v\big|_{y=0}}{x_v^o} + \frac{b}{RT}(x_v\big|_{y=0} - x_v^o)\right) - 4F N_v\big|_{y=0} LWR_W, \tag{S1.9}$$

In the present analysis, we only consider quasi-steady solutions for a purely sinusoidal voltage perturbation of radial frequency $\omega$ and voltage amplitude $\tilde{v}$:

$$V = -\tilde{v}\cos(\omega t) \tag{S1.10}$$

Substituting Eq. 1.12, S1.1, and S1.2 into S1.11, and nondimensionalizing the PDE and boundary conditions according to the dimensional scaling factors summarized in Table S1, we obtain:

$$A_o \frac{\partial \psi}{\partial \tau} = \frac{\partial}{\partial \xi}\left[\left(A_o + (A_o - 1)\psi\right)\frac{\partial \psi}{\partial \xi}\right] - \kappa\left[\left(1+\psi\right)^2 - e^{-2(A_o-1)\psi}\right] \tag{S1.11}$$

$$-\alpha\cos(\sigma\tau) = -\ln(1+\psi|_{\xi=0}) - (A_o - 1)\psi|_{\xi=0} + \gamma'\left[\left(A_o + (A_o-1)\psi|_{\xi=0}\right)\frac{\partial \psi}{\partial \xi}\bigg|_{\xi=0}\right] \tag{S1.12}$$

$$y\big|_{x\to\infty} = 0 \tag{S1.13}$$

where $\alpha = 2F\tilde{v}/RT$ and $\sigma = \tilde{\omega}L^2/A_o D_v$ are the dimensionless perturbation amplitude and frequency, respectively. The two remaining dimensionless groupings, $\gamma'$ and $\kappa$, are given by:

$$\kappa = \frac{2\mathfrak{R}_o^o L}{D_v c_O x_v^o}, \tag{S1.14}$$

which is a rate coefficient ratio for surface $O_2$ exchange vs. bulk diffusion, and

$$\gamma' = R_\Omega \frac{8F^2 c_O D_v x_v^o W}{RT}, \tag{S1.15}$$

which is the ratio of ohmic resistance in the electrolyte to a characteristic resistance for bulk diffusion.

For small, steady-periodic voltage perturbations (of the type used in EIS), Eqs. S1.15-1.17 can be linearized, allowing $y$ to be separated into time and spatial components of the form:[12]

$$\psi(\xi,\tau,\sigma) = \tfrac{\alpha}{2}\left[\bar{\psi}(\xi,\sigma)e^{j\sigma\tau} + \bar{\psi}^*(\xi,\sigma)e^{-j\sigma\tau}\right], \tag{S1.16}$$

where $\bar{\psi}(\xi,\sigma)$ (and its complex conjugate $\bar{\psi}^*(\xi,\sigma)$), is a complex stationary function expressing the real (in phase) and imaginary (out of phase) components of the oxygen vacancy displacement relative to the phase of the voltage perturbation. Linearizing (S1.11)-(S1.12) and substituting (S1.16) yields a 2nd order ODE and boundary conditions:

$$\frac{\partial^2 \bar{\psi}}{\partial \xi^2} = (2\kappa + j\sigma)\bar{\psi} \tag{S1.17}$$

$$\gamma' \frac{\partial \bar{\psi}}{\partial \xi}\bigg|_{\xi=0} = \bar{\psi}|_{\xi=0} - \frac{1}{A_o} \tag{S1.18}$$

$$\bar{\psi}|_{\xi\to\infty} = 0 \tag{S1.19}$$

This system of equations yields the solution

$$\bar{\psi} = \frac{1}{A_o} \frac{e^{-\xi\sqrt{2\kappa+j\sigma}}}{\left(1+\gamma'\sqrt{2\kappa+j\sigma}\right)}, \quad (S1.20)$$

Eq. (S1.20) has the same form as Eq. 1 in the main text. However, by convention we usually report phase relative to a positive cathodic polarization (which is opposite in sign to (S1.20)). Also, since the absorption coefficient $\beta$ relating $\psi$ to $\chi$ is empirical, we drop the $1/A_0$ scaling factor. Finally, Eq. 1.24 reveals that the diffusion-based scaling factors in Table S1 (chosen for mathematical convenience) are not the natural scaling for this 1-D solution. A more natural choice is to rescale $\gamma$ and $\omega$ to the Gerischer time and length scales defined in Eq. 2, which are related to $\kappa$ and $t^*$ in Table S1 by:

$$\sqrt{2\kappa} = \sqrt{t^*/t_G} = L/l_\delta = \gamma/\gamma' \quad (S1.21)$$

where $\gamma$ is the ratio of the electrolyte resistance to $R_G$. Substitution of these modifications into S1.24 yields Eq. 1.

With this model, the impedance is given by $Z = \tilde{v}/\bar{I}$, where $\bar{I}$ is the Fourier transform of the current, obtained by substituting (S1.18) and (S1.20) into the linearized equation for $I$, (S1.7), yielding:

$$\bar{I} = 4c_O F D_v x_v^o W \frac{\sqrt{2k+js}}{1+g\ell\sqrt{2k+js}} ae^{jst}, \quad (S1.22)$$

and thus:

$$Z = \frac{RT}{8F^2} \frac{1}{c_O D_v x_v^o W} \left(\gamma' + \frac{1}{\sqrt{2\kappa+j\sigma}}\right) \quad (S1.23)$$

Rescaling according to the relationships (S1.22), we arrive at Eq. 5 in the main text.

Figure S2 shows the vacancy concentration displacement profiles and impedance predicted by this model. The concentration profiles evolve from an exponential decay to a Warburg profile with increasing frequency, as discussed in the main text. The corresponding impedance shifts from the low-frequency, semicircular shape to a 45° tail over the same frequency range. These predictions are presented to complement Fig. 3 of the main text by illustrating the transition between limiting behaviors, and reinforce the relationship between vacancy profiles and impedance.

**Table S1 |** Relationships between the dimensional and dimensionless forms quantities in the 1-D model.

| Quantity | Dimensional Variable | Dimensionless Form | Relationship | Dimensional Group |
|---|---|---|---|---|
| Distance | $y$ | $\xi$ | $\xi = y/L$ | $L \equiv$ Film Thickness |
| Time | $t$ | $\tau$ | $\tau = \dfrac{t}{t^*}$ | $t^* = \dfrac{L^2}{A_o D_v}$ |
| Vacancy Concentration | $x_v$ | $\psi$ | $\psi = \dfrac{x_v - x_v^o}{x_v^o}$ | $x_v^o$ |
| Frequency | $\tilde{\omega}$ | $\sigma$ | $\sigma = \tilde{\omega} t^*$ | $t^* = \dfrac{L^2}{A_o D_v}$ |
| Voltage | $\tilde{v}$ | $\alpha$ | $\alpha = \dfrac{\tilde{v}}{V^*}$ | $V^* = \dfrac{RT}{2F}$ |
| Current | $i$ | $\lambda$ | $\lambda = \dfrac{i}{i^*}$ | $i^* = \dfrac{2F c_o D_v x_v^o}{L}$ |
| Resistance | $R_\Omega$ | $\gamma'$ | $\gamma' = \dfrac{R_\Omega}{R^*}$ | $R^* = \dfrac{RT}{8F^2 c_o D_v x_v^o W}$ |

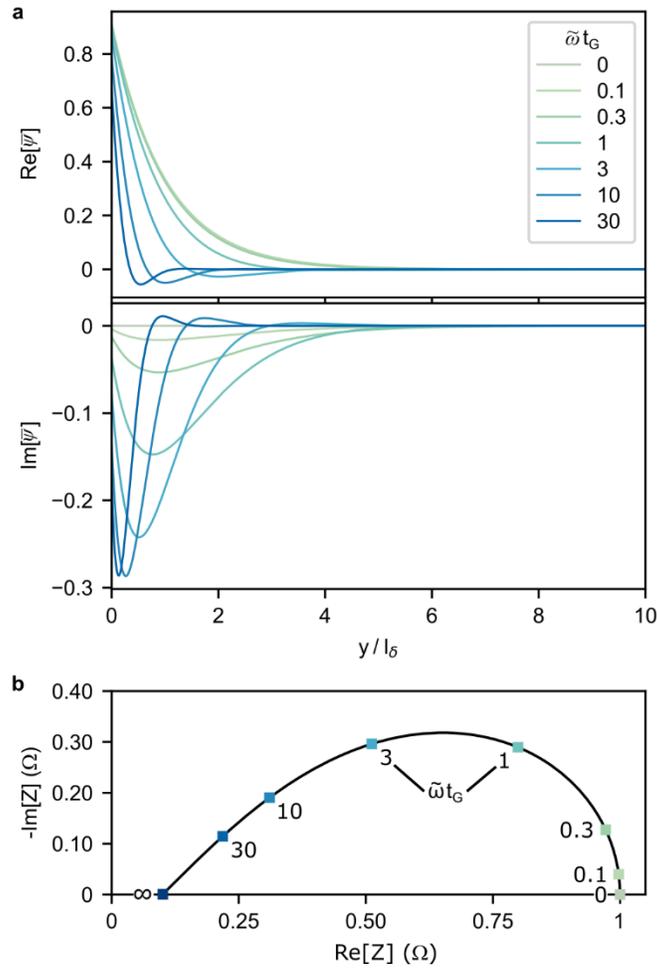

**Figure S2|** Predictions from the 1-D model with $A_o = 1$, $R_\Omega = 0.1$, and $R_G = 0.9$ a) vacancy concentration displacement profiles for varying $\tilde{\omega}$ at fixed $t_G$. b) Impedance, where the squares correspond to vacancy profiles above at the same frequency.

## Equivalent Circuit Model

Fig. 4c of the main text shows a diagram of the equivalent circuit model used to fit the measured impedance (not including electrolyte resistance). The bottom branch represents the impedance $Z_{NH}$ of the imperfectly insulating mask layer (which includes chemical capacitance of the LSC film above the mask. The top branch includes a series resistance ($R_c$) caused by interfacial ionic resistance, including constriction effects as ionic current funnels from the bulk electrolyte into the narrow gate area. The element ($Z_G$) corresponds to the Gerischer portion of Eq. 5, predicted from FR-XAS and described in the main text. The remaining parallel elements ($R_{gate}$ and $C_{gate}$) represent the surface reaction rate and chemical capacitance, respectively, of LSC directly above the gate region (which is not included in our 1-D model for $Z_G$).

Table S2 summarizes the parameter values used in the equivalent circuit model, which were estimated by a combination of a priori prediction and area-normalized measurements. $Z_G$ was calculated as already described and shown in Fig. 4a. We assumed LSC above the gate region is diffusionally equilibrated, with $O_2$ exchange limited by surface kinetics, analogously to unpatterned PLD films deposited directly onto the electrolyte. Using previous models for similarly prepared films,[3,6,13,14] we predicted the area specific resistance (ASR) and volume specific capacitance (VSC) as:

$$\text{ASR} = \frac{RT}{16F^2 \mathfrak{R}_O^o} \tag{S1.24}$$

$$\text{VSC} = \frac{4F^2 c_O}{V_m \left( c_{fit} + \frac{RT}{\delta^o} \right)} \tag{S1.25}$$

where $V_m$ is the molar volume (33.7 $\frac{cm^3}{mol}$) and $\delta$ is the nonstoichiometry, predicted using the same thermodynamic model used in predicting $Z_G$. The extensive quantities $R_{gate}$ and $C_{gate}$ were determined by scaling 1.10 and 1.11 to the known gate area and volume.

The value of $Z_{NH}(W)$ was taken from impedance measurements on a cell with a mask extending under the entire LSC film (no gates), to characterize the insulating properties of the mask (the subscript NH stands for "no holes"). As shown in Figure S3, this cell had a finite admittance despite having no gates to allow direct ion transfer between LSC and ceria. The PLD conditions for fabricating the "no hole" sample were similar to those used for the patterned electrode samples; however, the superficial area and film thicknesses differed. Thus we scaled $Z_{NH}$ assuming it is inversely proportional to mask area, and introduced an empirical scaling factor to account for differences in mask thickness, as discussed below.

**Table S2 |** Circuit element values for the model proposed in Fig. 4c of the main text.

| $pO_2$ (bar) | 0.01 | 0.1 | 1 |
|---|---|---|---|
| $R_c$ (Ω) | 68 | 68 | 68 |
| $C_{gate}$ (F) | 9.0 x 10$^{-5}$ | 7.5 x 10$^{-5}$ | 5.6 x 10$^{-5}$ |
| $R_{gate}$ (Ω) | 26,000 | 13,000 | 2600 |
| $R_G$ (Ω) | 1780 | 1630 | 1200 |
| $t_G$ (s) | 2.34 | 0.946 | 0.145 |

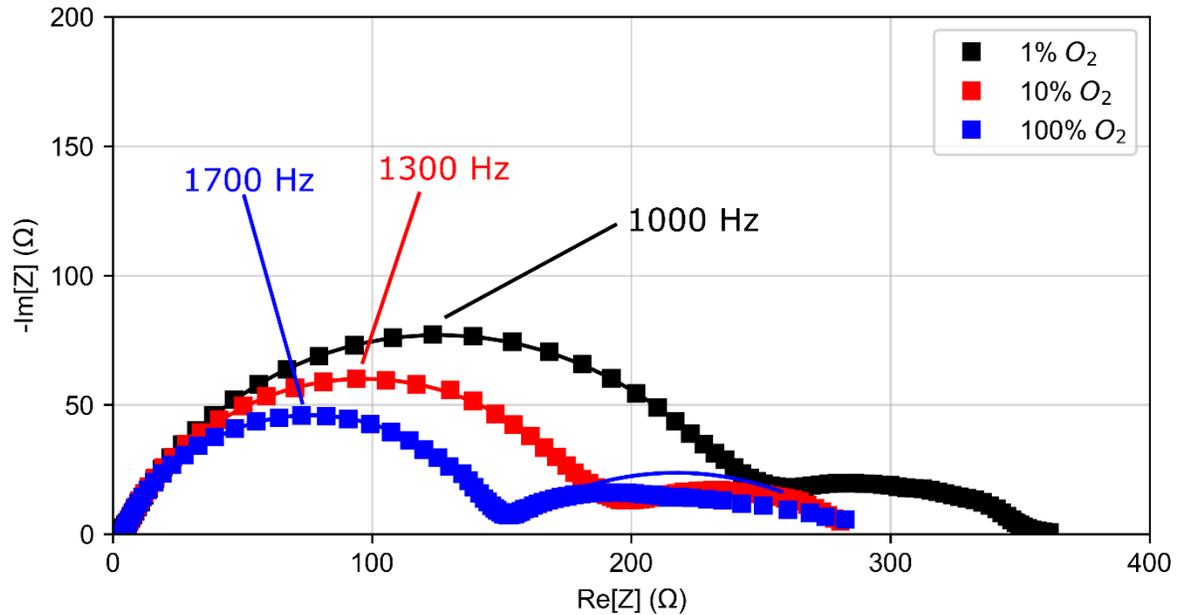

**Figure S3 |** Impedance spectra at 700 °C of the "no hole" LSC sample after subtracting the high-frequency intercept.

In order to account for possible inhomogeneities in cell geometry and physical properties, we introduced two independent empirical scaling factors in the impedance magnitude for the top and bottom branches of the equivalent circuit in Fig. 4c. The value of these scaling factors were applied across all datasets, and then optimized by least-squares fitting of the model to all the data (no other parameters were adjusted during the fit).

The net result of this calculation is shown in Fig. 4b, which corresponds to a top branch scaling factor of 2.1, and a bottom branch scaling factor of 0.70. In other words, the impedance predicted by the equivalent circuit model falls within a factor of 2 in magnitude and frequency of a priori expectations. It also accurately describes the *shape* of the impedance at high frequency (where $Z_G$ acting alone is expected to exhibit a 45° Warburg tail), as well as dispersion seen at the lowest frequencies.

As of the writing of this paper, we can only speculate why the scaling factors described above are not 1. It is important to remember that our predictions of the overall impedance are based on image data observed at a single location on the cell. Any variations in properties along the gates could easily result in average behavior differing from the model. Likely variations include variations in film and mask geometry, and inhomogeneous film properties. In particular Sr doping in LSC, which has a strong impact on most film properties, is known to vary across the thickness of a PLD film,[3,14] and could also easily vary with lateral position. Meanwhile, nonuniform film and mask thickness would impact $C_{gate}$, $Z_G$ and $Z_{NH}$. Notably our predictions of $Z_{NH}$ are based on measurements from a separate sample where the insulating layer was two times thicker than in the patterned electrode. If the mask insulation scales with this thickness, it would partly explain the observed 0.7 scaling factor for the lower circuit branch. However, we don't currently know enough about the origin of the leak current to draw a firm conclusion. Our current efforts include developing a more insulating mask to eliminate this issue as a factor. We are also developing 2D transport model for the film to more explicitly model the behavior at the highest frequencies.

# SI References